\documentclass[twocolumn]{aastex631}

\usepackage{amsmath,amsthm,amssymb,amsfonts}
\makeatletter
\renewcommand{\maketag@@@}[1]{\hbox{\m@th\normalsize\normalfont#1}}%
\makeatother

\makeatletter
\newcommand*\fsize{\f@size pt\relax}
\makeatother
\usepackage{comment}
\newcommand{\kepler}[0]{\emph{Kepler}}
\newcommand{\corot}[0]{\emph{CoRoT}}
\newcommand{\tess}[0]{\emph{TESS}}
\newcommand{\ktwo}[0]{\emph{K2}}

\newcommand{\gaia}[0]{\emph{Gaia}}
\newcommand{\nyquist}[0]{\emph{Nyquist}}
\newcommand{\teff}[0]{$T_{\text{eff}}$}
\newcommand{\rsolar}[0]{$R_{\sun}$}

\newcommand{\logg}[0]{$\log g$}
\newcommand{\Dnu}[0]{$\Delta\nu$}
\newcommand{\numax}[0]{$\nu_{\rm max}$}

\begin{document}

\title{Detection of Solar-like Oscillations in Sub-giant and Red Giant Stars Using 2-minute Cadence TESS Data}

\shorttitle{Detection of Solar-like oscillations by TESS}
\shortauthors{Zhou et al.}

% \correspondingauthor{Shaolan Bi}
\email{bsl@bnu.edu.cn}
% \correspondingauthor{Jie Yu}
\email{jie.yu@h-its.org}

\author[0009-0004-9024-9666]{Jianzhao Zhou}
\affiliation{Institute for Frontiers in Astronomy and Astrophysics,
Beijing Normal University,
Beijing 102206, China}
\affiliation{Department of Astronomy,
Beijing Normal University, Beijing 100875,
People's Republic of China}

\author[0000-0002-7642-7583]{Shaolan Bi}
\affiliation{Institute for Frontiers in Astronomy and Astrophysics,
Beijing Normal University,
Beijing 102206, China}
\affiliation{Department of Astronomy,
Beijing Normal University, Beijing 100875,
People's Republic of China}

\author[0000-0002-0007-6211]{Jie Yu}
\affiliation{Max Planck Institute for Solar System Research, Justus-von-Liebig-Weg 3,
D-37077 Gottingen, Germany}

\author[0000-0003-3020-4437]{Yaguang Li}
\affiliation{Sydney Institute for Astronomy (SIfA),
School of Physics, University of Sydney,
NSW 2006, Australia}
\affiliation{Stellar Astrophysics Centre,
Department of Physics and Astronomy,
Aarhus University, Ny Munkegade 120, DK-8000 Aarhus C, Denmark}

\author[0000-0002-3672-2166]{Xianfei Zhang}
\affiliation{Institute for Frontiers in Astronomy and Astrophysics,
Beijing Normal University,
Beijing 102206, China}
\affiliation{Department of Astronomy,
Beijing Normal University, Beijing 100875,
People's Republic of China}

\author[0000-0001-6396-2563]{Tanda Li}
\affiliation{Institute for Frontiers in Astronomy and Astrophysics,
Beijing Normal University,
Beijing 102206, China}
\affiliation{Department of Astronomy, Beijing Normal University,
Beijing 100875, People's Republic of China}
\affiliation{School of Physics and Astronomy,
University of Birmingham, Edgbaston, Birmingham B15 2TT, UK}

\author[0000-0003-2908-1492]{Liu Long}
\affiliation{Institute for Frontiers in Astronomy and Astrophysics,
Beijing Normal University,
Beijing 102206, China}
\affiliation{Department of Astronomy,
Beijing Normal University, Beijing 100875,
People's Republic of China}

\author{Mengjie Li}
\affiliation{Institute for Frontiers in Astronomy and Astrophysics,
Beijing Normal University,
Beijing 102206, China}
\affiliation{Department of Astronomy,
Beijing Normal University, Beijing 100875,
People's Republic of China}

\author[0000-0003-0795-4854]{Tiancheng Sun}
\affiliation{Institute for Frontiers in Astronomy and Astrophysics,
Beijing Normal University,
Beijing 102206, China}
\affiliation{Department of Astronomy,
Beijing Normal University, Beijing 100875,
People's Republic of China}

\author[0009-0009-1338-1045]{Lifei Ye}
\affiliation{Institute for Frontiers in Astronomy and Astrophysics,
Beijing Normal University,
Beijing 102206, China}
\affiliation{Department of Astronomy,
Beijing Normal University, Beijing 100875,
People's Republic of China}

\begin{abstract}

Based on all 2-minute cadence \tess{} light curves from Sector 1 to 60, we provide a catalog of 8,651 solar-like oscillators, including frequency at maximum power (\numax{}, with its median precision, $\sigma$=5.39\%), large frequency separation (\Dnu{}, $\sigma$=6.22\%), seismically derived masses, radii and surface gravity. In this sample, we have detected 2,173 new oscillators and added 4,373 new \Dnu{} measurements. Our seismic parameters are consistent with those from \kepler{}, \ktwo{}, and previous \tess{} data.  The median fractional residual in \numax{} is $1.63\%$ with a scatter of $14.75\%$, and in \Dnu{} it is $0.11\%$ with a scatter of $10.76\%$.
We have detected 476 solar-like oscillators with \numax{} exceeding the \nyquist{} frequency of \kepler{} long-cadence data during the evolutionary phases of sub-giant and the base of the red-giant branch, which provide a valuable resource for understanding angular momentum transport.
\end{abstract}
\keywords{\href{http://astrothesaurus.org/uat/1583}{Asteroseismology (1583)}; \href{http://astrothesaurus.org/uat/481}{Subgiants (481)}; \href{http://astrothesaurus.org/uat/655}{Red giants (655)}; \href{http://astrothesaurus.org/uat/1234}{Lightcurves (1234)}} 

\section{Introduction} \label{sec:intro}

Astroseismology, the study of stellar oscillations, offers a powerful tool to infer stellar interiors \citep[e.g.,][]{Christensen-Dalsgaard1984,Aerts2010}. In recent decades, space missions such as \corot{} \citep{Baglin2006, Auvergne2009}, \kepler{} \citep{Borucki2010}, and \ktwo{} \citep{Howell2014} have provided long-duration, high-quality photometry data, leading to a revolution in the study of solar-like oscillations. These missions enabled the study of solar-like oscillations in hundreds of main-sequence and sub-giant stars \citep[e.g.,][]{Chaplin2011, Chaplin2014, Li2020, Mathur2022a}, as well as tens of thousands of red giants \citep[e.g.,][]{Hekker2011, Stello2013, Huber2014, Mathur2016, Yu2016, Yu_2018, Hon2019}, revealing new aspects of stellar structure and evolution \citep[e.g.,][]{Hekker2017,Aerts2019}. 

The NASA Transiting Exoplanet Survey Satellite (\tess{}) mission \citep{Ricker2015} has provided an opportunity to study solar-like oscillations in stars across the entire sky. Previous studies had extensively used \tess{} data to characterize solar-like oscillators, with a primary focus on the Continuous Viewing Zones (CVZs) near the ecliptic pole due to longest observation duration \citep{SilvaAguirre2020a,Mackereth2021, Hon2022, stello2022}. \cite{Hon2021} initially used deep learning techniques to detect solar-like oscillations in red giants across the full sky on the first two years of \tess{} full-frame images (FFI, Sector 1 to 26), identifying about 158,000 giants with \numax{}. \cite{Hatt2023} detected 4,177 solar-like oscillators using both 2-minute and 20-second cadence data (Sector 1 to 46), reporting \numax{} and \Dnu{} estimates. 

By the end of Sector 60, \tess{} had completed observations of both the north and south ecliptic hemispheres for the second time. More than half of targets with 2-minute data had been observed in at least two sectors. Longer photometric time series provide higher frequency resolution in the Fourier domain, leading to improved precision in asteroseismic measurements. In this work, we aim to perform a complete search for solar-like oscillators and provide their global seismic parameters using \tess{} 2-minute cadence data. In addition, by combining \teff{} from the \gaia{} DR3 Radial Velocity Spectrometer (RVS) survey \citep{Recio-Blanco2023}, we estimate stellar radii, masses and surface gravity using the seismic scaling relations.

\section{Data Selection} \label{sec:data}

\subsection{Preliminary data selection}\label{sec:pre-selected}

We download all available 2-minute cadence light curves spanning Sector 1 to 60 from the Mikulski Archive for Space Telescopes (MAST)\footnote{Data can be found in MAST: \dataset[10.17909/t9-nmc8-f686]{http://dx.doi.org/10.17909/t9-nmc8-f686}}. These light curves were extracted and de-trended by the \tess{} Science Processing Operations Center (SPOC) pipeline \citep{Twicken2016,Jenkins2017}.

\begin{figure}
% \plotone{DataSelection.png}
\includegraphics[height=11cm,width=8cm]{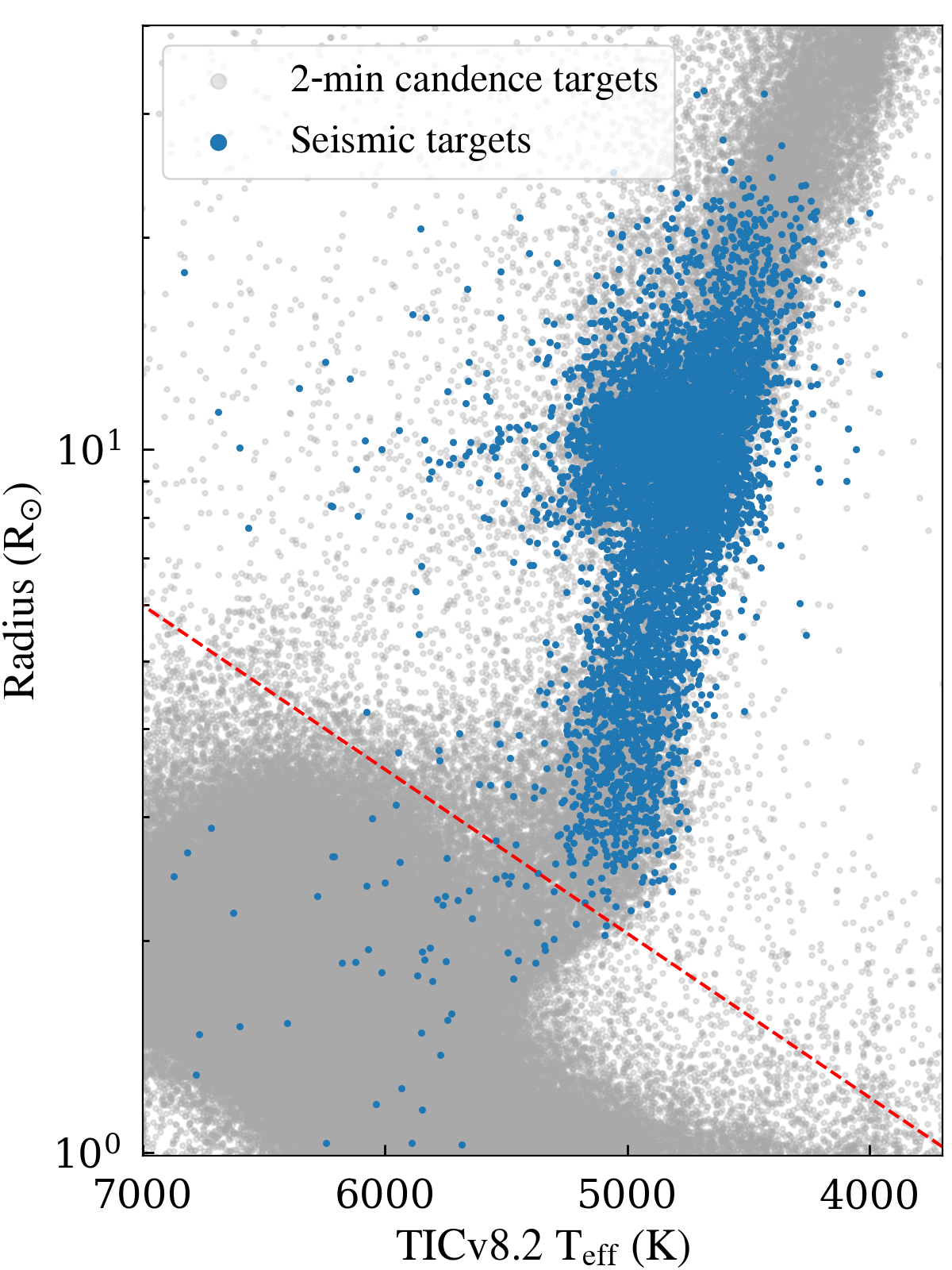}
\caption{$\rm{R}$ versus \teff{} for 2-minute cadence targets and seismic targets. The \teff{} and radii are sourced from TICv8.2. Light gray dots show the 2-minute cadence targets, blue dots show the seismic targets, and the red dashed line, reference from \citet{Hon2019}, shows the boundary used to distinguish early sub-giant stars from those ascending the red giant branch.}
\label{fig:DataSelection}
\end{figure}

A sample of stars selected for oscillation detection is based on the effective temperature (\teff) and radius ($\rm{R}$) values from the \tess{} Input Catalog version 8.2 \citep[TICv8.2;][]{Stassun2019}, with criteria of $1$\rsolar$\le R \le 40$ \rsolar\ and $3700$K $\le$ \teff $\le$ $7000$K. This resulted in the identification of 193,020 candidates, depicted as gray dots in Figure \ref{fig:DataSelection}.
To categorize the stars into dwarfs and giants, we employ the relation $R=10^{p}$\rsolar, with $p=\left( \frac{T_{\text{eff}}(K)}{7000} - \frac{3}{7} \right) \log\left(\frac{300}{7}\right)+\log(0.7)$ proposed by \citet{Hon2019} as the boundary (indicated by the red dashed line in Figure \ref{fig:DataSelection}). This categorization results in 154,817 main sequence stars and subgiants, as well as 38,203 red giants.

\begin{figure*}
    \centering{
    \includegraphics[scale=0.3]{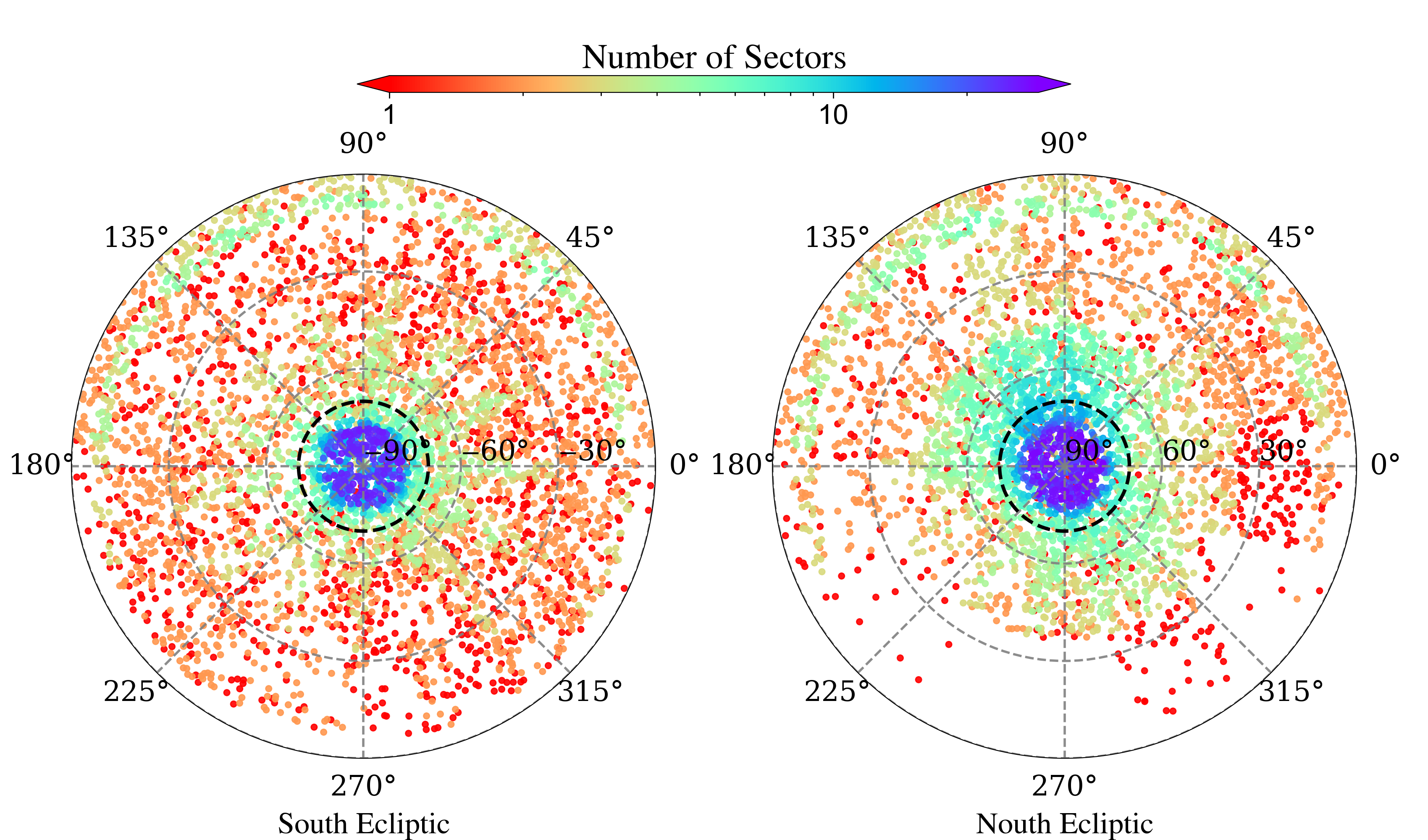}
    \caption{The distribution of detected solar-like oscillators across the celestial sphere. The color bar corresponds to the observation duration, denoted by the number of sectors. Some stars (green dots) are observed for 2--3 sectors near the ecliptic because \tess{} focused on a specific segment of the ecliptic from Sector 42 to 46 coinciding with the portion of the \ktwo{} observation zones. The northern ecliptic hemisphere contains gaps to avoid excessive contamination from stray Earth- and moonlight, and the corresponding region appears at low latitudes close to the ecliptic. The black dashed circles represent the CVZs.\label{fig:detection_map}}
    }
\end{figure*}

\subsection{Seismic data detection\label{sec:detection}}

We transform all the light curves (PDCSAP data) of our pre-selected stars (Section \ref{sec:pre-selected}) into power density spectra using the Lomb-Scargle periodogram method \citep{VanderPlas2018}. We apply a 5 $\sigma$-clipping to remove outliers, and divide the light curves by a 10-day median filter to eliminate low-frequency signals in each sector, and concatenate the light curves of all available sectors \citep{Garcia2011}.

We detect the oscillation power excess in each power density spectrum using the collapsed autocorrelation function (collapsed ACF) method \citep{Huber2009a}. First, we divide the power spectrum in equally logarithmic bins and smooth the result using an empirical $40\%$ percentile filter to obtain a crude estimate of the background. Second, the residual power spectrum, obtained from dividing the power density spectrum by the estimated background, is segmented into overlapping subsets. The width of each subset is approximately 4$\Delta\nu_{\rm{exp}}$ around its central frequency ($\nu_{\rm{center}}$), where $\Delta\nu_{\rm{exp}}$ is estimated as $0.263 \times \nu_{\rm{center}}^{0.772}$ \citep{Stello2009}. For each subset, we calculate the absolute ACF and then collapse the ACF by its referring $\nu_{\rm{center}}$. Finally, we smooth the collapsed ACFs with an empirical 7 $\mu$Hz filter and fit them with a Gaussian profile centered on their maximum peak, along with constant noise.

We retain stars with a signal-to-noise ratio(SNR) greater than 1.5, and identify 7,870 solar-like oscillators. For stars with $1.2 \leq \text{SNR} \leq 1.5$, we carefully visually inspect their power density spectra for the presence of power excess, and confirm 209 oscillators. The 8,080 solar-like oscillators are shown in Figure \ref{fig:DataSelection} (blue dots), including 61 main sequence stars and subgiants, and 8,019 red giants. This indicates a solar-like oscillation detection rate of approximately $20\%$ in red giants. Furthermore, we repeat the same procedure for stars not presented in TICv8.2 and identify an additional 571 oscillators. In total, we have identified an asteroseismic sample of 8,651 stars.

Figure \ref{fig:detection_map} illustrates the distribution of the sample across the ecliptic celestial sphere, covering nearly the entire sky. \tess{} focused on a specific segment of the ecliptic during the fourth year, spanning from Sector 42 to 46, to coincide with the portion of the \ktwo{} observation zones. Consequently, some stars (green dots) were observed for 2-3 sectors near the ecliptic. In order to minimize the contamination from stray Earth- and moonlight, \tess{} boresights toward a latitude of $+85\arcdeg$ in some sectors, which leads to an incomplete coverage of the northern hemisphere at low latitudes. The black dashed circles within $20\arcdeg$ of the northern and southern ecliptic poles represent the CVZs.

\section{Measuring Global Seismic Parameters} \label{sec:method}
\subsection{Measuring the frequency at maximum power}
\begin{figure}[ht]
\centering
\includegraphics[height=8.98cm,width=7.15cm]{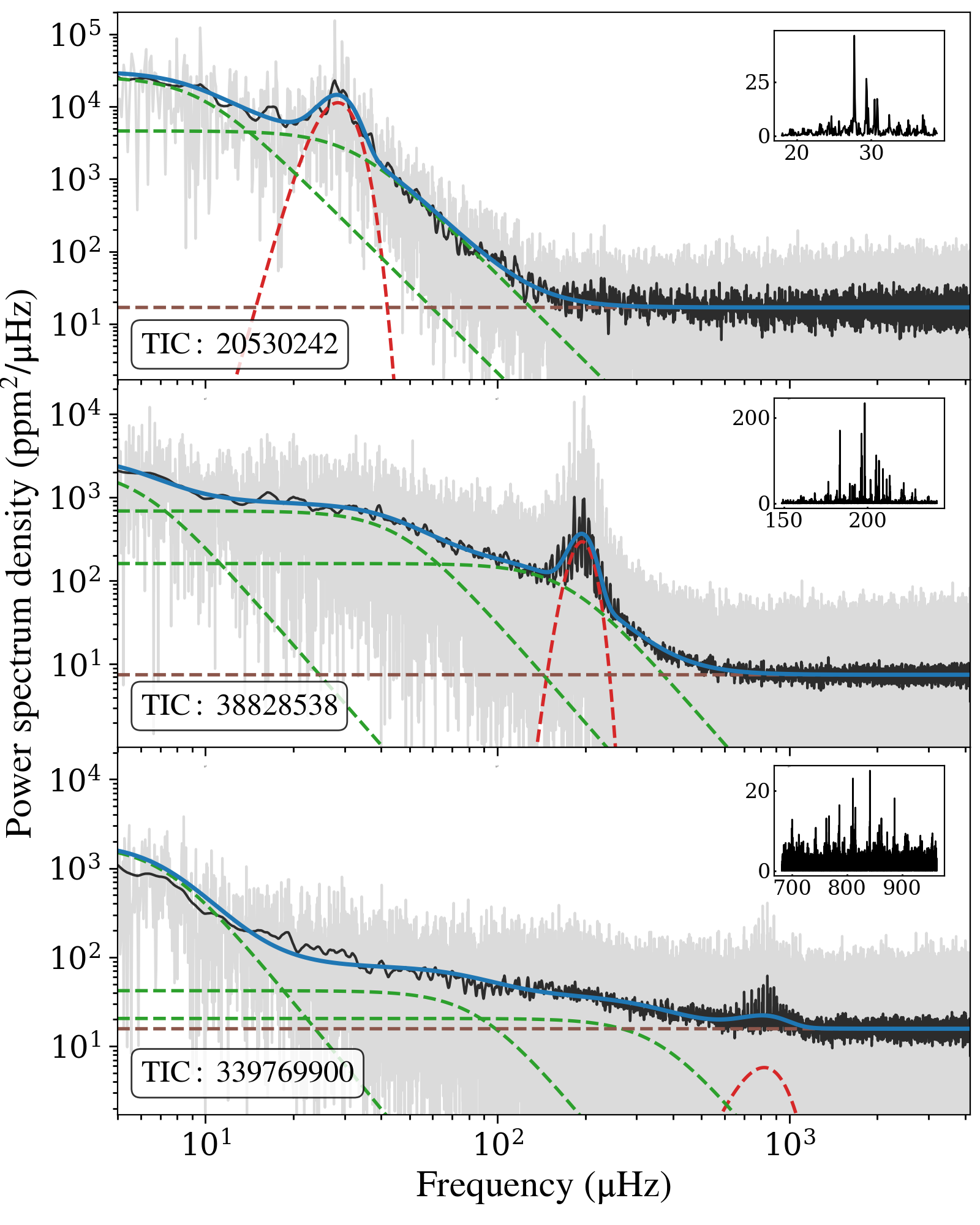}
\caption{Representative power spectra of \tess{} 2-min cadence light curves.
In each panel, the gray line shows the real data, and the black line shows the data smoothed by a boxcar filter of 3 $\mu$Hz wide.
The solid blue line shows the MCMC fitting result, with the red dashed line for the fitted Gaussian envelope, the green dashed curve for the three Harvey components, and the brown straight line for the white noise.  The inset figure shows the power spectrum near \numax{} divided by the background. }
\label{fig:fit_mcmc}
\end{figure}

\begin{figure*}[ht]
{\centering
\includegraphics[height=13.7cm,width=16cm]{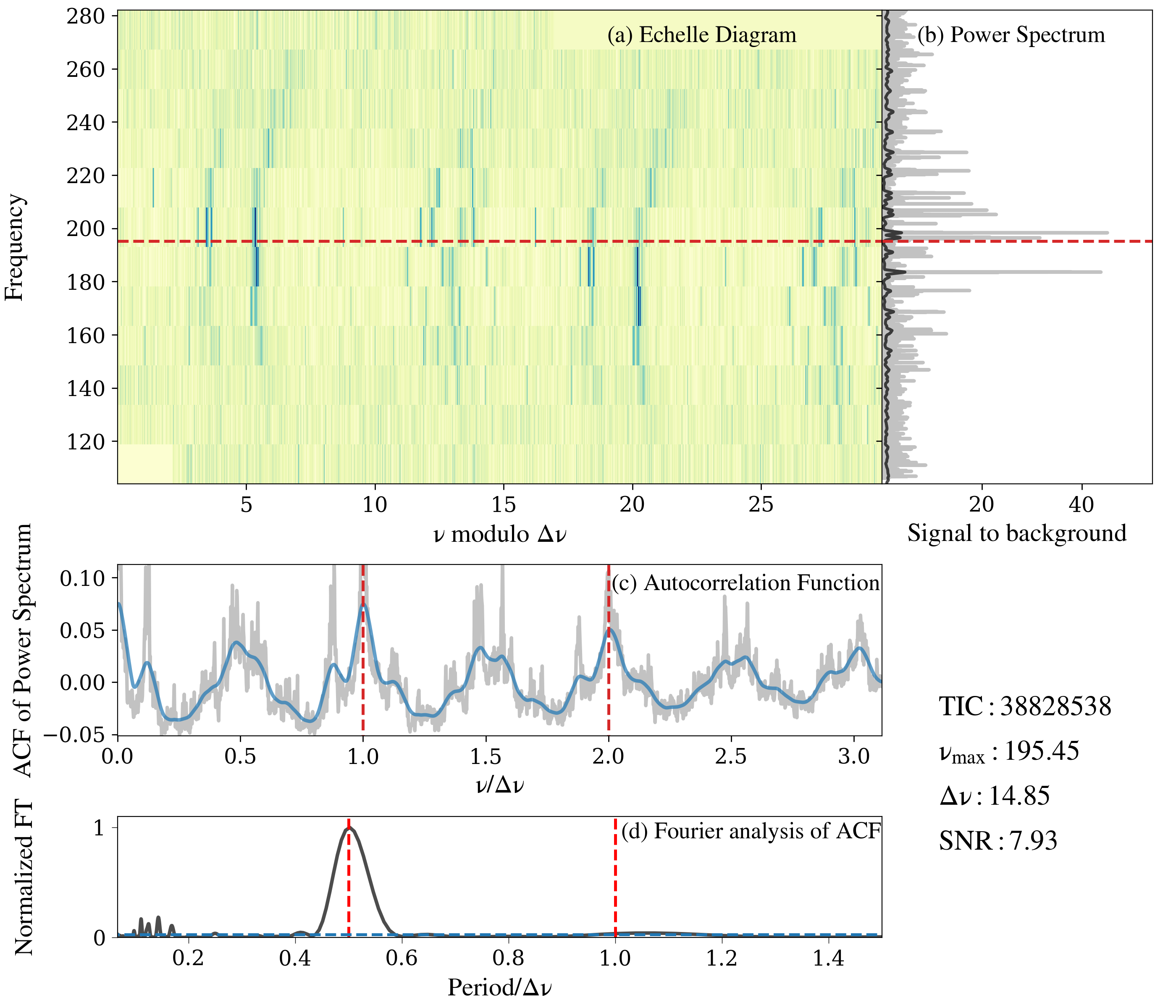}
\caption{An example of measuring \Dnu{} for TIC 38828538 : (a) the \'echelle diagram, the red dashed line represents the value of \numax{}; (b) the residual power spectrum;  (c) the autocorrelation function, the two red dashed lines represent the positions of \Dnu{} and twice \Dnu{} from left to right; (d) the Fourier transforms of ACF, the two red dashed lines represent the positions of half of \Dnu{} and \Dnu{} in period from left to right, the blue dashed line shows noise. \label{fig:Diagnostic plot}}
}
\end{figure*}

We use the center of the fitted Gaussian profile to the collapsed ACF (Section \ref{sec:detection}) as the initial $\nu_{\rm{max, guess}}$. To fit the power density spectrum, we employ a model that consists of a Gaussian envelope, three background Harvey components, and white noise \citep{Chaplin2014}:
\begin{equation}
    \begin{aligned}
    &P(\nu) = \\
    &\eta(\nu)^{2}\left[\sum_{i}^{3} \frac{2\sqrt{2} a_{i} / \pi b_{i}}{1+\left(\nu / b_{i}\right)^{4}}+H_{g}\right. 
    \left.\exp \frac{-\left(\nu-\nu_{\text {max }}\right)^{2}}{2 \sigma^{2}}\right] + W_{\text {n}} ,\label{fit:model}
    \end{aligned}
\end{equation}
where $\eta(v)=\rm{sinc}\left(\pi \nu/2\nu_{\rm{nyq}}\right)$ accounts for the frequency-dependent attenuation resulted from the observational signal discretization, and $\nu_{\rm{nyq}}$ is the \nyquist{} frequency \citep[e.g.,][]{Chaplin2011b,Kallinger2014}. For other parameters, $a_i$ and $b_i$ represent the root-mean-square (rms) and the characteristic frequency of the $i$th Harvey component, respectively. $H_g$, $\nu_{\text{max}}$, and $\sigma$ are the height, the central frequency, and the width of the Gaussian envelope, respectively. $W_{\text{n}}$ corresponds to the contribution of white noise.

\begin{table*}[ht!]
\tabcolsep 0.4truecm
\centering{
\caption{Stellar Global Oscillation Parameters\label{tab:asteroseismic_parameters}}
\begin{tabular}{lccccccccc}
\hline
\hline
TIC    & $\rm{Tmag}$ & $N_{\rm sectors}$  & \numax    & $\sigma\left(\nu_{\rm max}\right)$ & \Dnu  & $\sigma\left(\Delta\nu\right)$ & $\rm{SNR}$& $\rm{Types}$&$\rm{Source}$\\
% \hline
         &  mag &   & $\mu$Hz   & $\mu$Hz & $\mu$Hz     & $\mu$Hz   \\   
\hline
1608&8.78&2&43.66&5.17&4.67&0.33&7.64&--&--\\
13727&7.31&1&188.99&5.04&16.91&0.94&7.39&--&--\\
80047&9.76&1&64.07&5.60&7.81&0.99&6.36&Binary&4\\
89696&8.63&1&43.13&1.84&4.81&0.57&3.71&--&--\\
92094&8.21&2&44.17&2.51&--&--&2.36&--&--\\
99433&4.42&2&73.55&5.20&--&--&2.82&--&--\\
105245&8.26&2&272.89&7.94&18.16&0.96&3.53&--&--\\
$\cdots$&$\cdots$&$\cdots$&$\cdots$&$\cdots$&$\cdots$&$\cdots$&$\cdots$&$\cdots$&$\cdots$\\
471011913&6.38&2&256.66&3.33&18.59&0.27&6.17&--&--\\
900749927&5.35&4&39.69&1.91&3.84&0.34&6.87&--&--\\
\hline
% \caption{Stellar global parameters of \tess{} targets presenting solar-like oscillations }
\end{tabular}
}
\tablecomments{The source of the adopted stellar types for each star is indicated by the following: (1) Spectroscopic Binary Orbits Ninth Catalog, (2) the \tess{} eclipsing binary catalog, (3) NASA's Exoplanet Archive, (4) \gaia{} DR3 \texttt{nss\_two\_body\_orbit} Catalog. SNR indicates the signal-to-noise ratio of \Dnu{}. The machine-readable table is fully accessible.}
\end{table*}

We estimate \numax{} and its uncertainty by Bayesian inference using Monte Carlo Markov Chain (MCMC) simulations \citep{Foreman-Mackey2013}. The initial fitting parameters for MCMC are from the Maximum Likelihood Estimation (MLE) method \footnote{The initial fitting parameters used in MLE are derived from $\nu_{\rm{max, guess}}$.} \citep{Huber2009a, Kallinger2010}.
The minimum number of steps for the MCMC estimation is 3000 and the maximum number of steps is 5000. The \numax{} is estimated as the median of the posterior probability distribution, and the uncertainties are approximated as the $16^{th}/84^{th}$ percentiles. Representative examples of the background fitting are shown in Figure \ref{fig:fit_mcmc}.

\subsection{Measuring the large frequency separation}

We use the ACF method to measure \Dnu{} values \citep{Huber2009a, Chontos2022}, shown in Figure \ref{fig:Diagnostic plot}. Figure \ref{fig:Diagnostic plot}(a) displays the \'echelle diagram of the oscillation modes. To prepare for the \Dnu{} measurement, we first normalize the power density spectrum by dividing it by the MCMC-fitted background (Gaussian component excluded). Then, we restrict the normalized power density spectrum to the frequency range of \numax{}$\pm 3\Delta\nu_{\rm{exp}}$, as shown in Figure \ref{fig:Diagnostic plot} (b). Within this frequency range, we calculate the ACF and apply a boxcar filter with an empirical width of $0.2\Delta\nu_{\rm{exp}}$. Finally, \Dnu{} is measured as the maximum value within the range of $0.7$--$1.3\ \Delta\nu_{\rm{exp}}$ in the smoothed ACF, as depicted in Figure \ref{fig:Diagnostic plot}(c).

To evaluate the significance of the \Dnu{} measurements, we calculate SNR through dividing the maximum value of normalized Fourier transforms (FT) on the ACF by the noise, corresponding to the half of \Dnu{} or \Dnu{}. The noise is represented by the rms of the normalized FT on the ACF, as shown in Figure \ref{fig:Diagnostic plot}(d). Consequently, we obtain a sample of 7,509 stars with valid \Dnu{} measurements, adopting only \Dnu{} measurements with SNR $\ge 3$.

Following \citet{Huber2011}, the uncertainties of \Dnu{} are estimated by conducting 500 perturbations on the power-density spectrum using a $\chi^2$ distribution with two degrees of freedom. For each perturbation, the fitting procedure is repeated, and the standard deviation of 500 measurements is considered as the uncertainty.

\section{Results} \label{sec:result}

\subsection{Asteroseismic Sample\label{sec:AsteroseismicSample}} 

We present an asteroseismic sample of 8,651 solar-like oscillators with \numax{}, including 7,509 stars with \Dnu{}. Notably, 2,173 stars from this sample are new oscillators that were not previously detected \citep{Hon2021,Hon2022,Hatt2023}. Compared to \citet{Hatt2023}, we add 4,373 new \Dnu{} of stars. 
Additionally, we flag 781 binaries and 85 exoplanet host stars by cross-matching the sample with Spectroscopic Binary Orbits Ninth Catalog\footnote{\url{https://sb9.astro.ulb.ac.be/}}, NASA Exoplanet Archive\footnote{\url{https://exoplanetarchive.ipac.caltech.edu/}}, the \tess{} Eclipsing Binary Catalog\footnote{\url{http://tessebs.villanova.edu/}}, and \gaia{} DR3 \texttt{nss\_two\_body\_orbit} Catalog, respectively \citep{Pourbaix2004,Howard2022,Prsa2022,GaiaCollaboration2022}.
The results are listed in Table \ref{tab:asteroseismic_parameters}.

\begin{figure}
\centering{
\includegraphics[height=6.5cm,width=8cm]{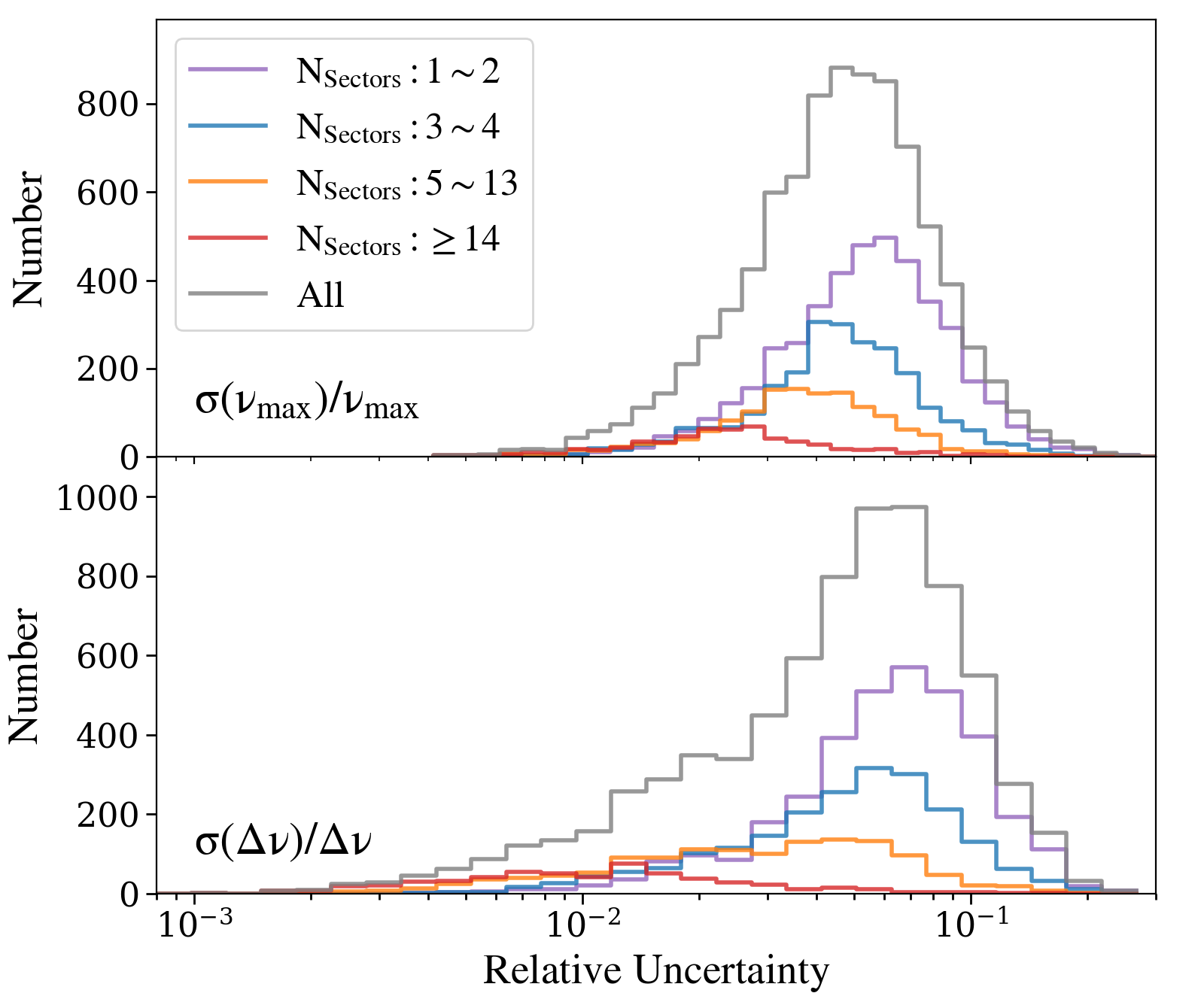}
\caption{Histogram of relative uncertainties for \numax{} and \Dnu{}. $N_{\rm{sectors}}$ shows the number of sectors. The gray line represents the entire sample, while the purple, blue, yellow, and red lines represent stars observed for 1 to 2 sectors, 3 to 4 sectors, 5 to 13 sectors, and stars observed for more than 14 sectors, respectively. \label{fig:uncertainties} }
}
\end{figure}

Figure \ref{fig:uncertainties} shows the histogram of relative uncertainties for \numax{} and \Dnu{}. The number of sectors ($N_{\rm{sectors}}$) marks the observation duration, with longer duration corresponding to lower uncertainties. This indicates that longer duration significantly improves measurement precision.

\begin{figure}
\centering{
    \includegraphics[height=4.5cm,width=8.5cm]{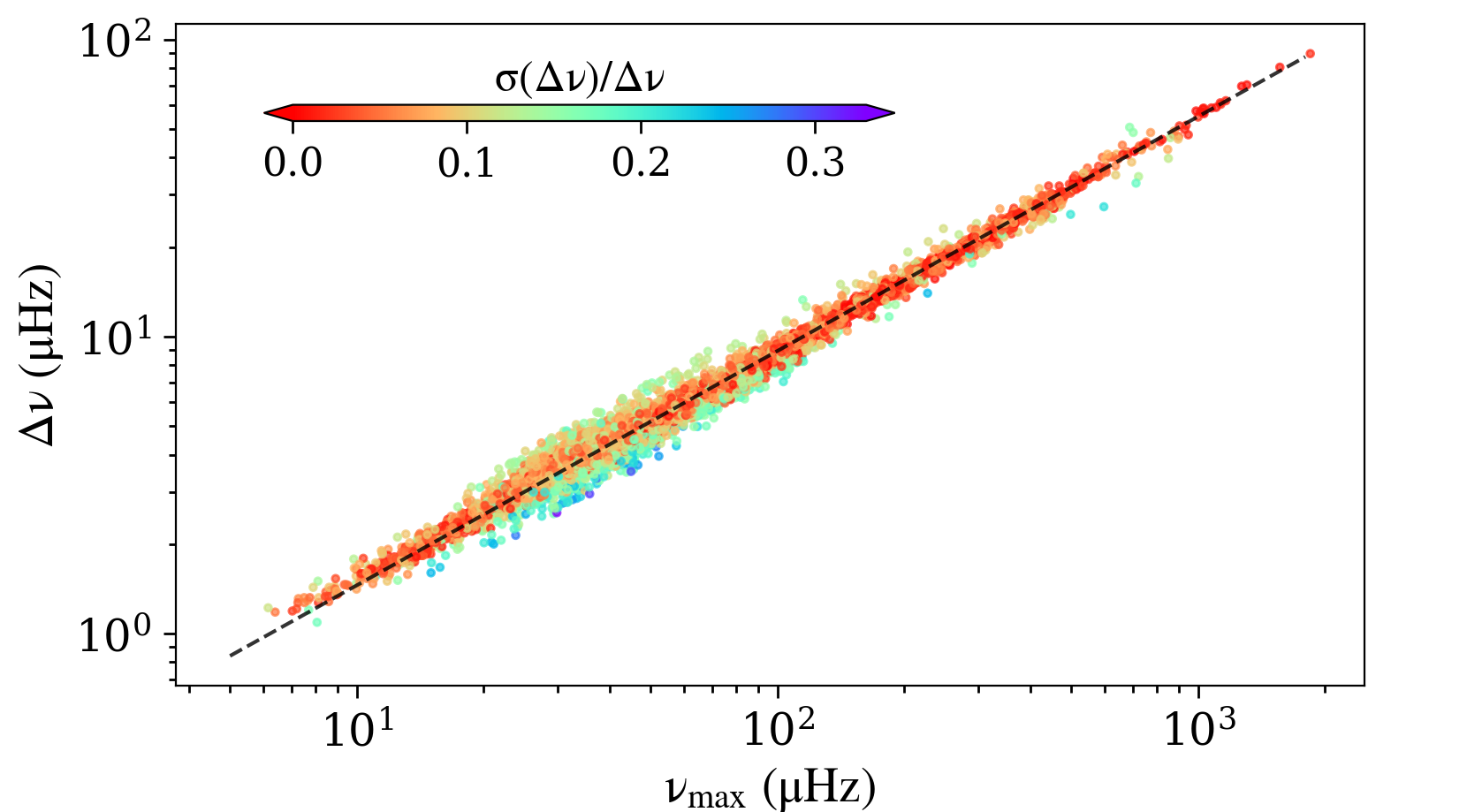}
    \caption{The relation of \numax{} and \Dnu{} in logarithmic coordinates. The color bar shows the relative error of the measured \Dnu{}. The black dotted line follows a MCMC fitted power-law relation: $\Delta\nu =\alpha \cdot (\nu_{\rm{max}})^\beta$, where $\alpha=0.236\pm0.001$ and $\beta=0.789\pm0.001$.\label{fig:result}}
}
\end{figure}

Figure \ref{fig:result} shows the well-established power-law relation of \numax{} and \Dnu{} \citep{Stello2009,Hekker2009}. The black dotted line is expressed as $\Delta\nu =\alpha \cdot (\nu_{\rm{max}})^\beta$. It is fitted by an MCMC method, and consequently, we obtain $\alpha=0.236\pm0.001$ and $\beta=0.789\pm0.001$ . 
The \numax{} and \Dnu{} exhibit a linear relation in logarithmic coordinates, especially for stars with $\sigma(\Delta\nu)/\Delta\nu$ $\leq$ 0.1. 
The stars having \Dnu{} precision worse than 0.1 and $\nu_{\rm{max}} \sim 20\mbox{--}100\mu \rm{Hz}$, may correspond to red clump (RC), as they exhibit complex power spectra \citep{Hon2017}.

\begin{figure}
    \centering
    \includegraphics[height=4.1cm,width=8.6cm]{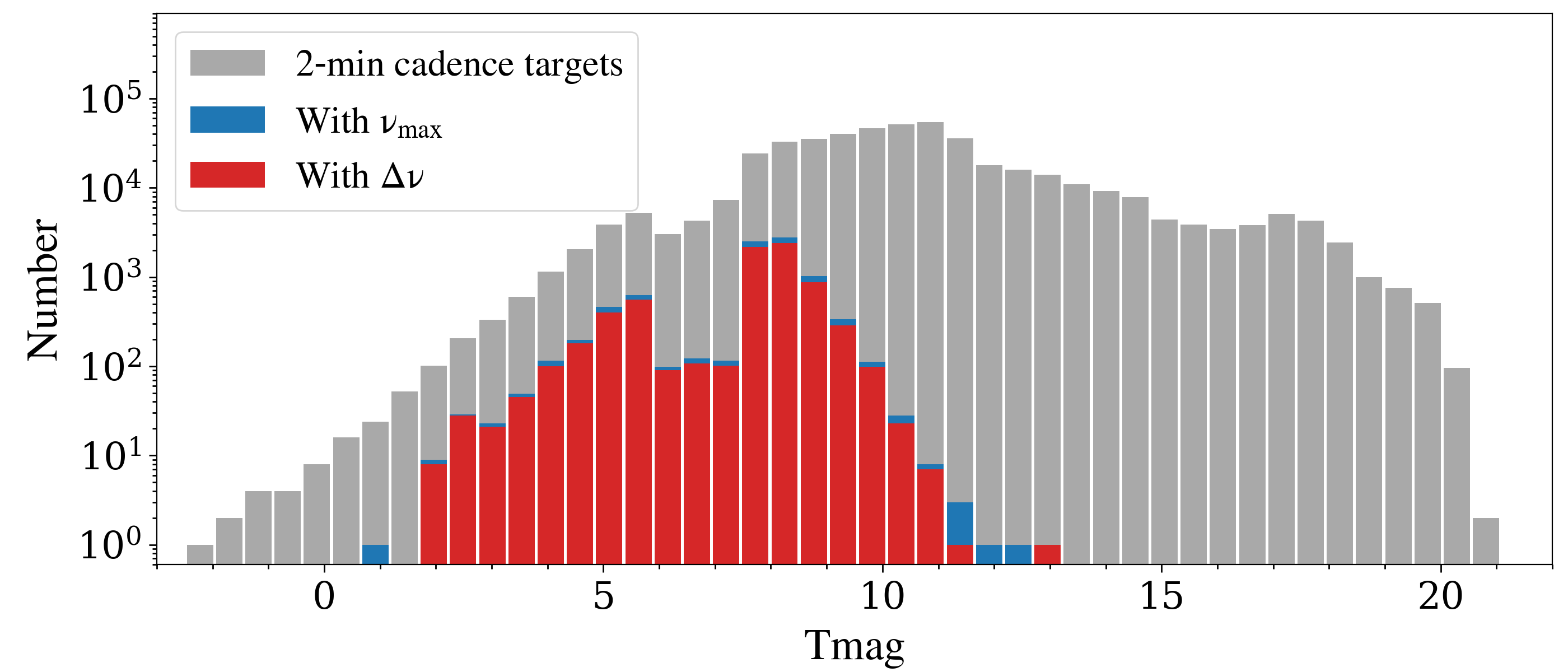}
    \caption{Histogram of \tess{} magnitude ($\rm{Tmag}$) for all 2-minute cadence targets and the seismic stars. The gray color shows all 2-min cadence targets, the blue color shows oscillators with \numax{}, and the red color shows stars with \Dnu{}. 
    % The bottom panel shows the number of stars versus the star magnitude. 
    \label{fig:distribution_all}
    }
\end{figure}

Figure \ref{fig:distribution_all} shows the histogram of \tess{} magnitude ($\rm{Tmag}$) for all 2-minute cadence stars and our sample. The number of our sample increases with $\rm{Tmag}$ at $\rm{Tmag}$ $<$ 5, which is consistent with the overall sample. However, for $\rm{Tmag}$ $>$ 9, the number of oscillators significantly decreases with increasing \tess{} magnitude. This suggests an optimal magnitude range of $6 \leq \rm{Tmag} \leq 9$ for observing solar-like oscillations. Additionally, there is a gap around $\rm{Tmag}=7$, consisting with \citet{Hatt2023}. The gap exists in both all 2-minute cadence targets and our sample. We examined the histogram of TESS magnitude for each sector from Sector 1 to 60 and found the gap in each of them. It possibly results from the inhomogeneous selection for TESS observations of 2-minute cadence data.

\subsection{Comparison of \numax{} and \Dnu{} Measurements}\label{sec:CompareWithPreviousLiterature}

\begin{figure*}[ht]
\centering
\includegraphics[width=0.43\linewidth]{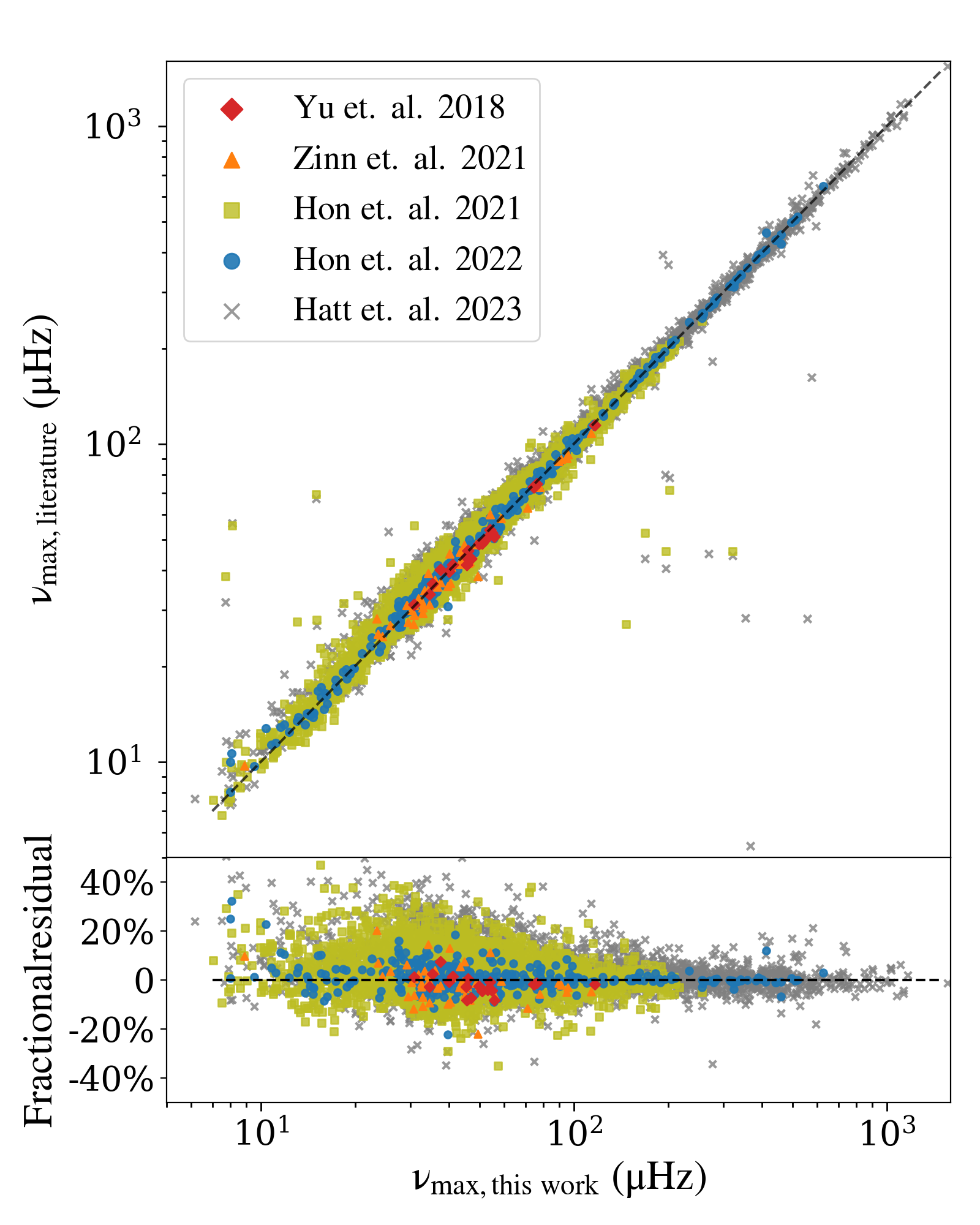}
\includegraphics[width=0.43\linewidth]{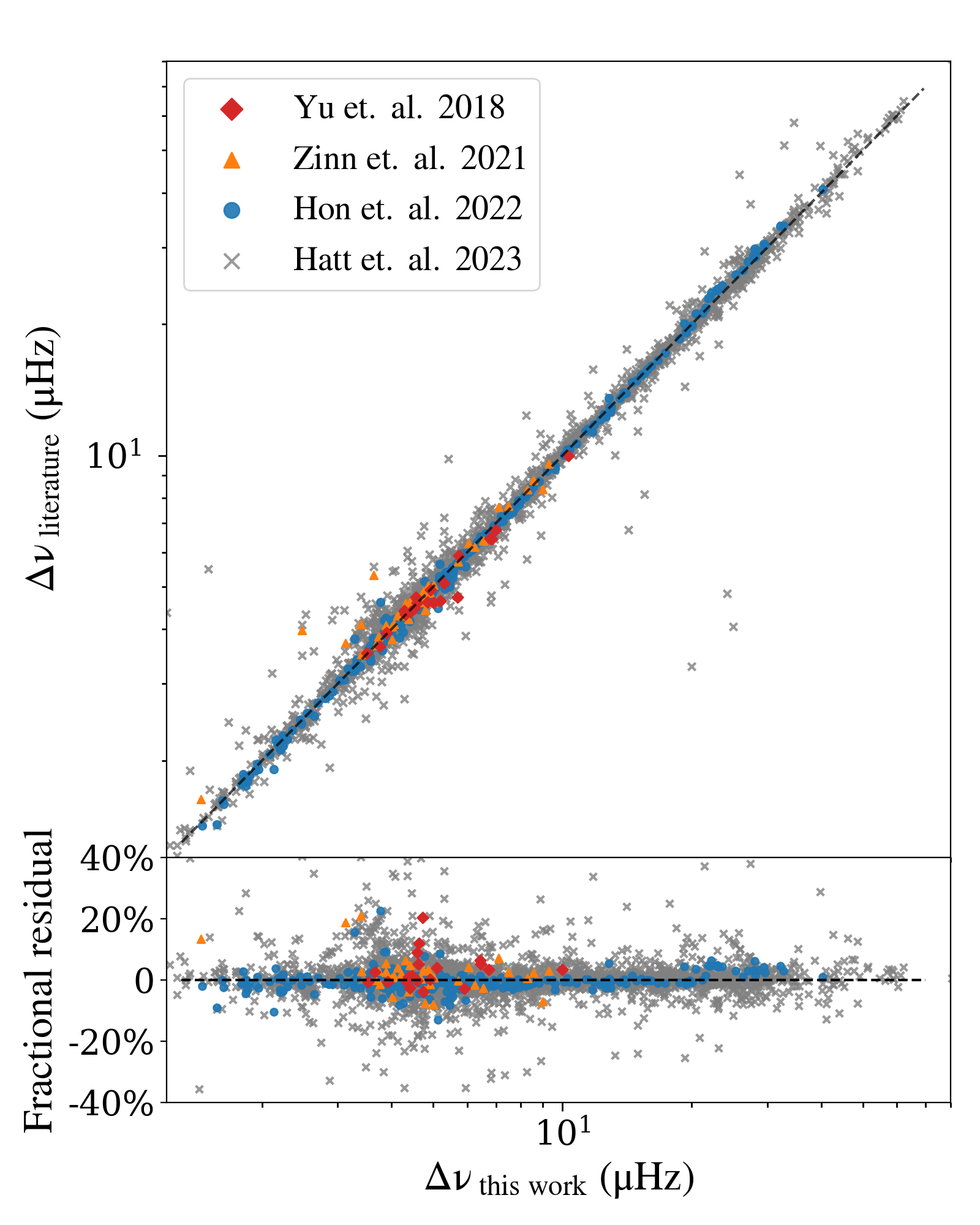}
\caption{Comparison between global seismic parameters measured in this work and those from previous literature. The black dashed lines in the top panel show the one-to-one relation between the two parameters. The bottom left panel shows the fractional residuals of $\nu_{\rm max}$, calculated as $\left(\nu_{\rm max,literature}-\nu_{\rm max,this\ work}\right)/\nu_{\rm max,this\ work}$. Similarly, the bottom right panel displays the fractional residuals of $\Delta\nu$, calculated as $\left(\Delta\nu_{\rm literature}-\Delta\nu_{\rm this\ work}\right)/\Delta\nu_{\rm this\ work}$.\label{fig:CompareWithOther}}
\end{figure*}

In Figure \ref{fig:CompareWithOther}, we compare our global seismic parameters with those of common stars from literature. The results demonstrate good agreement, despite the differences in the methods and data. The median fractional residual in \numax{} is $1.63\%$ with a scatter of $14.75\%$, while the median fractional residual in \Dnu{} is $0.11\%$ with a scatter of $10.76\%$, as shown in Table \ref{tab:comparison}.

\begin{figure*}
{\centering
\includegraphics[height=10.5cm,width=16cm]{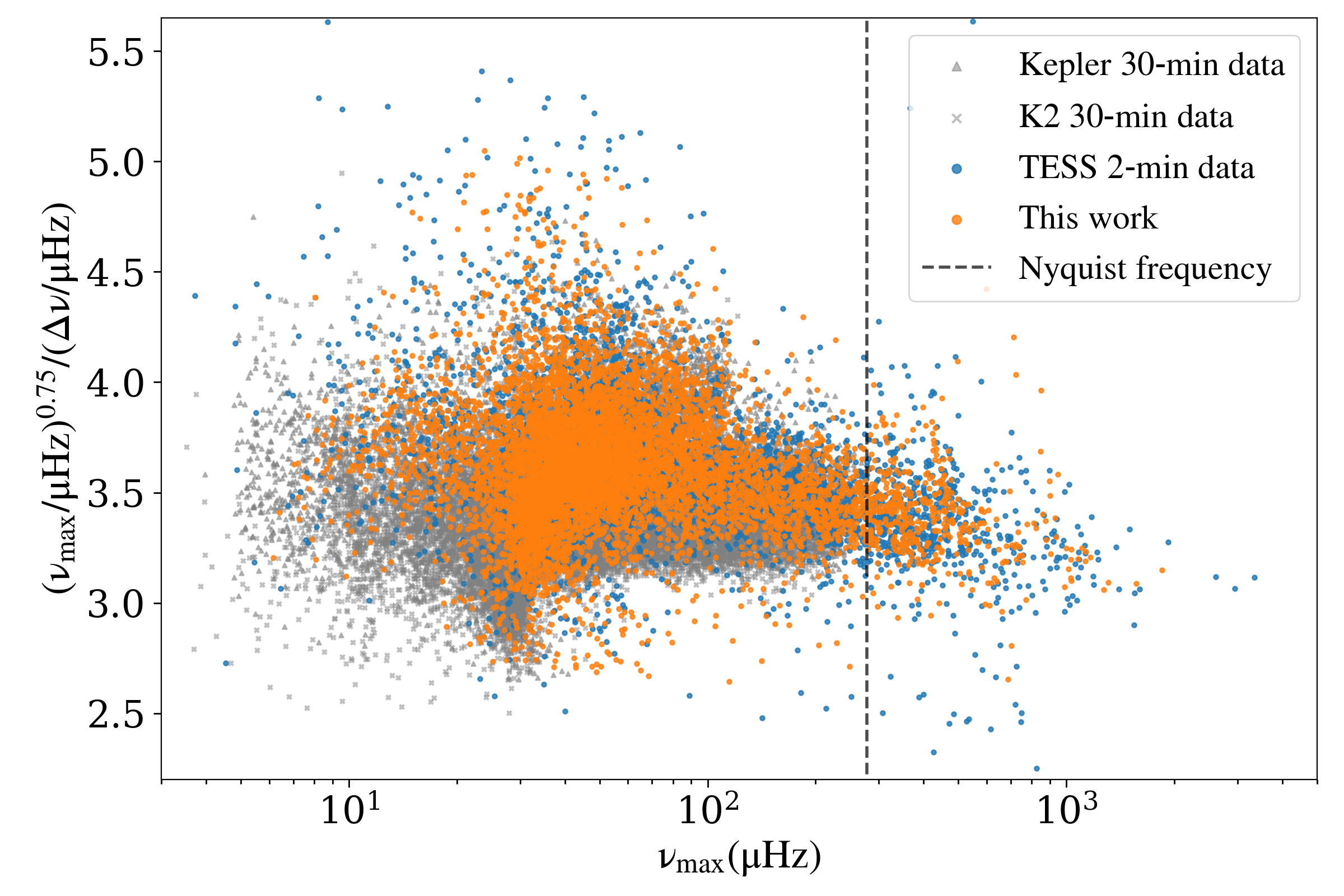}

\caption{\numax{} vs. $\nu_{\rm{max}}^{0.75}/\Delta\nu$ diagram. The horizontal axis shows \numax{}, while the vertical axis shows $\nu_{\rm max}^{0.75}/\Delta\nu$, which is a mass proxy related to temperature. The gray triangles and crosses represent samples from \kepler{} and \ktwo{} long-cadence data, respectively. Blue dots show sample from previous \tess{} 2-minute data as presented by \citet{Hatt2023} and yellow dots show our sample. \label{fig:SeismicMeasurements}}
}
\end{figure*}

Figure \ref{fig:SeismicMeasurements} shows the distribution in \numax{} - $\nu_{\rm{max}}^{0.75}/\Delta\nu$ for our sample and stars in \kepler{}/\ktwo{} long-cadence data and \tess{} 2-minute data from \citet{Hatt2023}. Compared to \kepler{}/\ktwo{} sample, on the one hand, there are fewer high-luminosity red giants in our sample, because our oscillators were observed within shorter duration. On the other hand, near the \nyquist{} frequency, it is possible to measure \Dnu{} with long-cadence data but challenging to measure \numax{}, because the granulation background fitting could be biased and \nyquist{} aliases may occur \citep{Yu2016,Yu_2018}. In this context, \tess{} 2-minute cadence data prove valuable. 

Notably, we have detected 401 solar-like oscillators with \numax{} exceeding the \nyquist{} frequency of \kepler{}/\ktwo{} long-cadence data. These oscillators are more-evolved subgiants or low-luminosity red giants, whose solar-like oscillation were seldom detected by either long- or short-cadence observation of the previous \kepler{}/\ktwo{} mission. Such stars transform from nearly-uniform rotation to differential rotation, helping us to understand angular momentum transport \citep[e.g.,][]{Aerts2019,Deheuvels2020,Kuszlewicz2023,Wilson2023}. 

\begin{table*}
\tabcolsep 0.6truecm
\centering{
\caption{Comparison of Global Seismic Parameters with Previous Literature}
\begin{tabular}{rccccc}
\hline
\hline
Missions: & \kepler{} & \ktwo{} &  & \tess{} & \\
\cline{4-6}
Cadences: & 30 mins  & 30 mins  & 30 mins  &30 mins  & 120 and 20 secs \\
\hline
Common Stars:&   $20^{1}$ &$ 51^{2} $  & $5375^{3}$  & $348^{4}$  & $3129^{5}$ \\
\hline
\numax{}\ Median Residual:& $-1.81\%$ & $-2.03\%$ & $1.37\%$ & $0.42\%$ & $2.49\%$ \\
        Scatter:&         $3.63\%$  & $7.41\%$  & $13.53\%$ & $5.05\%$ & $17.21\%$\\
\hline
\Dnu{}\ Median Residual:&   $-1.99\%$ & $2.44\%$  &    --    & $-0.34\%$& $0.11\%$\\
      Scatter:&           $4.88\%$  & $13.17\%$  &     -- & $3.68\%$ & $10.58\%$\\

\hline
\end{tabular}
\tablecomments{Stars of Common are sourced from: (1) \citet{Yu_2018}, (2) \citet{zinn2019}, (3) \citet{Hon2019}, (4) \citet{Hon2021}, (5) \citet{Hatt2023}.}
\label{tab:comparison}
}
\end{table*}

\subsection{Fundamental stellar parameters}

By cross-matching our sample to Gaia DR3 RVS spectra data, we obtain 7,173 stars with asteroseismic parameters and \teff{}. We estimate radius ($R_{\rm{seismic}}$), mass ($M_{\rm{seismic}}$), and surface gravity (\logg{}) of these stars by the scaling relations \citep{Ulrich1986,Brown1991,Kjeldsen-Bedding1995,Belkacem2011}:

\begin{equation}
\frac{M_{\rm{seismic}}}{M_{\sun}} \approx \left(\frac{\nu_{\rm max}}{\nu_{\rm max,\sun}}\right)^3 \left( \frac{\Delta\nu}{\Delta\nu_{\sun}}\right)^{-4} \left(\frac{T_{\text{eff}}}{T_{\text{eff},\sun}}\right)^{3/2},
\label{eq:M_sca}
\end{equation}
\begin{equation}
\frac{R_{\rm{seismic}}}{R_{\sun}} \approx \left(\frac{\nu_{\rm max}}{\nu_{\rm max,\sun}}\right) \left( \frac{\Delta\nu}{\Delta\nu_{\sun}}\right)^{-2} \left(\frac{T_{\text{eff}}}{T_{\text{eff},\sun}}\right)^{1/2},
\label{eq:R_sca}
\end{equation}
\begin{equation}
\frac{\rm{g}}{\rm{g_{\sun}}} \approx \left(\frac{\nu_{\rm max}}{\nu_{\rm max,\sun}}\right) \left(\frac{T_{\text{eff}}}{T_{\text{eff},\sun}}\right)^{1/2},
\label{eq:logg}
\end{equation}
where $\nu_{\rm max, \sun} = 3090$ $\mu$Hz, $\Delta\nu_{\sun} = 135.1$ $\mu$Hz, and $T_{\rm eff, \sun} = 5777$ K adopted from \citet{Huber2013}. 
The estimates of $M_{\rm{seismic}}$, $R_{\rm{seismic}}$, and \logg{} are listed in Table \ref{tab:fundamental_parameters}, with their median uncertainties of $9.20\%$, $6.24\%$ and 0.01 dex ($0.79\%$), respectively.

\begin{table*}
\tabcolsep 0.3truecm
\caption{Fundamental stellar parameters \label{tab:fundamental_parameters}}
\centering{
\begin{tabular}{lcccccc}
\hline
\hline
TIC    & \teff  & \logg{}  & $M_{\rm{seismic}}$  & $R_{\rm{seismic}}$ & $R_{\rm{SED}}$ & $L_{\rm{SED}}$\\ 
% \hline
       &   $K$ & $c.g.s$ & $M_{\sun}$   & $R_{\sun}$ & $R_{\sun}$ & $log(L/L_{\sun})$\\ 
\hline
1608      &4739.0  $\pm$ 11.0   &2.55 $\pm$ 0.03 &1.47 $\pm$ 0.14  &10.72 $\pm$ 0.19 &11.09 $\pm$ 0.11 & 1.747 $\pm$ 0.016\\
13727     &4795.0  $\pm$ 162.5  &3.19 $\pm$ 1.89 &0.71 $\pm$ 0.06  &\ 3.56 $\pm$ 0.23  &\ 4.02 $\pm$ 0.04 & 0.908 $\pm$ 0.040\\
80047     &4736.0  $\pm$ 28.5   &2.73 $\pm$ 0.09 &0.60 $\pm$ 0.14  &\ 5.62 $\pm$ 0.89  &\ 6.52 $\pm$ 0.24 & 1.282 $\pm$ 0.035\\
89696    &4768.0  $\pm$ 22.0  &2.55 $\pm$ 0.04 &1.27 $\pm$ 0.42  &\ 10.01 $\pm$ 1.88  &\ 8.81 $\pm$ 0.10 & 1.554 $\pm$ 0.016\\
$\cdots$&$\cdots$&$\cdots$&$\cdots$&$\cdots$&$\cdots$&$\cdots$\\
11688264  &4756.0  $\pm$ 10.0   &2.20 $\pm$ 0.01 &0.69 $\pm$ 0.32  &10.95 $\pm$ 2.87  &11.72 $\pm$ 0.11 & 1.801 $\pm$ 0.016\\
11738052  &4854.0  $\pm$ 3.5    &2.78 $\pm$ 0.02 &1.12 $\pm$ 0.25  &7.15  $\pm$ 1.04  &\ 8.58 $\pm$ 0.04 & 1.562 $\pm$ 0.015\\
12063724  &4720.0  $\pm$ 11.5   &2.56 $\pm$ 0.03 &1.70 $\pm$ 0.20  &11.40 $\pm$ 0.08  &10.01 $\pm$ 0.07 & 1.651 $\pm$ 0.015\\
12333486  &4677.0  $\pm$ \ 8.5   &2.43 $\pm$ 0.02 &1.82 $\pm$ 0.44  &13.70 $\pm$ 1.93  &13.08 $\pm$ 0.35 & 1.867 $\pm$ 0.027\\
12358786  &4747.0  $\pm$ 10.0   &2.69 $\pm$ 0.03 &1.04 $\pm$ 0.08  &\ 7.70 $\pm$ 0.54  &\ 7.97 $\pm$ 0.07 & 1.464 $\pm$ 0.016\\
12376694  &4595.0  $\pm$ \ 9.0   &2.37 $\pm$ 0.02 &0.71 $\pm$ 0.03  &\ 9.14 $\pm$ 0.40  &10.77 $\pm$ 0.09 & 1.666 $\pm$ 0.015\\
$\cdots$&$\cdots$&$\cdots$&$\cdots$&$\cdots$&$\cdots$&$\cdots$\\
\hline
% \caption{Stellar global parameters of \tess{} targets presenting solar-like oscillations }
\end{tabular}
}
\tablecomments{Catalog of fundamental stellar parameters for 7,173 stars. \teff{} are collected from \gaia{} DR3 RVS spectra, the stellar $\rm{M_{\rm{seismic}}}$,$R_{\rm{seismic}}$ and \logg{} are provided by scaling relations, $\rm{R_{SED}}$ and $\rm{L_{SED}}$ obtained through the SED fitting. The machine-readable table is fully accessible.}
\end{table*}

\begin{figure}[ht]
    \includegraphics[height=7.2cm,width=8.6cm]{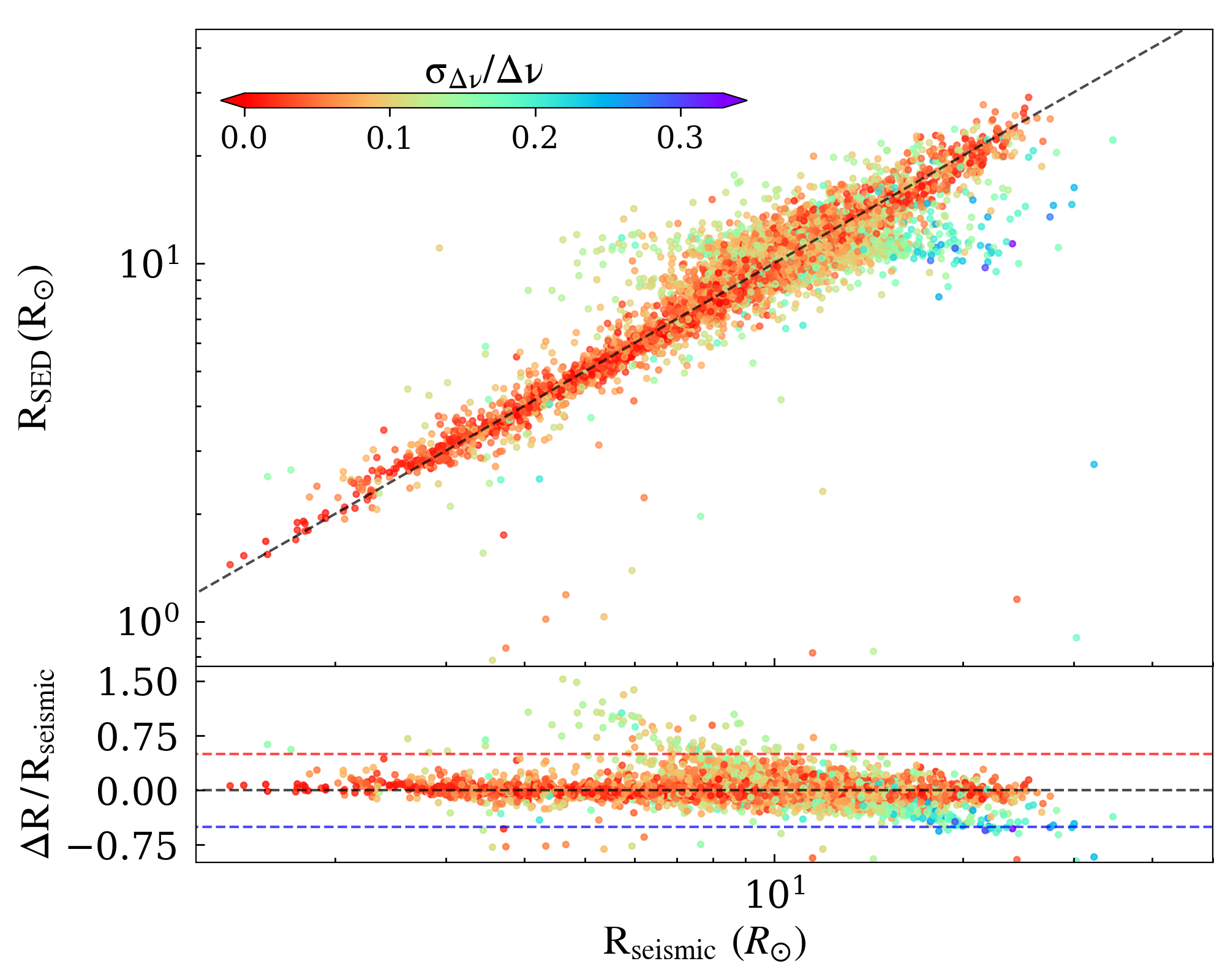}
    \caption{Comparisons of radii from the asteroseismic scaling relations with radii from the SED fitting. The color bar represents the relative error of \Dnu{}. The bottom panel displays fractional residuals between radius estimates, denoted as $\Delta R = R_{\rm{SED}} - R_{\rm{seismic}}$.\label{fig:Rcompare}}
\end{figure}
To validate our asteroseimic radii, we use another independent method to derive radii and luminosities for these stars. We employ the SEDEX pipeline \citep{Yu2021,Yu2023} alongside MARCS model spectra for performing the spectral energy distribution (SED) fitting. Our approach adopts spectroscopic \teff{}, \logg{}, and $\rm{[M/H]}$ priors (from Gaia DR3 RVS spectra) and combines them with apparent magnitudes from 32 band passes across nine photometric databases \citep{Yu2023} to derive both extinction and bolometric fluxes. Leveraging \gaia{} DR3 parallaxes, we then compute luminosities and deduce stellar radii in conjunction with the spectroscopic \teff{}. The uncertainties in bolometric fluxes are assessed through a Bayesian framework, and the uncertainties in luminosities and radii were determined via error propagation given \teff{} uncertainties. It is noted that the above two methods adopt the same effective temperature, implying that the measurements in both approaches may be subject to potential systematic effects related to \teff{}. Figure \ref{fig:Rcompare} shows the comparison between $R_{\rm{seimic}}$ and $R_{\rm{SED}}$. The result reveals a good agreement, with a median fractional residual of $-0.79\%$ and a standard deviation of $16.60\%$. This consistency is partly attributed to employing the same effective temperature in both measurements. The estimates of $R_{\rm{SED}}$ and $L_{\rm{SED}}$ are also listed in Table \ref{tab:fundamental_parameters}.

\section{Conclusions} \label{sec:conclusions}

We have presented a sample of 8,651 solar-like oscillators with \numax{} measurements, including 7,509 stars with \Dnu{} using \tess{} 2-min cadence light curves. Comparing with literature, we have newly detected 2,173 oscillators and added 4,373 \Dnu{} measurements. Our seismic parameters demonstrate good consistency with those from previous studies. 
The median fractional residual for \numax{} is $1.63\%$ with a scatter of $14.75\%$, and the median fractional residual for $\Delta \nu$ is $0.11\%$ with a scatter of $10.76\%$.

We have detected 476 solar-like oscillators that exhibit \numax{} values exceeding the \nyquist{} frequency of \kepler{}/\ktwo{} long-cadence data, which increases the sample size of more-evolved subgiants and low-luminosity red giants. Such oscillators may provide observational constraints on the stellar internal rotation profiles, which potentially contributes to our understanding of angular momentum transport.

We have estimated asteroseismic masses (with a median precision of $9.21\%$), radii (with a median precision of $6.24\%$), and \logg{} for a subset of 7,173 stars cross-matched from Gaia DR3 RVS spectra data. Our asteroseismic radii are in good agreement with the radii from the SED fitting.

Our sample covers the entire sky, showing the advantage of the \tess{} mission to detect solar-like oscillators. With further observations by \tess{}, a greater number and diversity of potential solar-like oscillators are expected to be detected. This will provide valuable observational targets for future missions, such as PLATO \citep[to be launched in 2026;][]{rauerPLATOMission2014} , which will significantly improve observations of stars with detectable solar-like oscillations. A higher-precision sample of solar-like oscillators spanning from main sequence stars to red giants, will provide new perspectives on stellar structure and evolution.

\section*{Acknowledgement}
% \begin{acknowledgments}
We thank Timothy R. Bedding, Ruijie Shi and HuanYu Teng for helpful discussions. This work is supported by the Joint Research Fund in Astronomy (U2031203) under cooperative agreement between the National Natural Science Foundation of China (NSFC) and Chinese Academy of Sciences (CAS), and NSFC grants (12090040, 12090042, 12073006). 
We also acknowledge the science research grant from the China Manned Space Proiectwith No.CMS-CSST-2021-A10, and CSST project. 

This paper includes data collected by the TESS mission. Funding for the TESS mission is provided by the NASA Explorer Program. Funding for the TESS Asteroseismic Science Operations Centre is provided by the Danish National Research Foundation (Grant agreement no.: DNRF106), ESA PRODEX (PEA 4000119301) and Stellar Astrophysics Centre (SAC) at Aarhus University. We acknowledge the use of public TESS data from pipelines a the TESS Science Office and at the TESS Science Processing Operations Center. Resources supporting this work were provided by the NASA High-End Computing (HEC) Program through the NASA Advanced Supercomputing (NAS) Division at Ames Research Center for the production of the SPOC data products.

This work has made use of data from the European Space Agency (ESA) mission {\it Gaia} (\url{https://www.cosmos.esa.int/gaia}), processed by the {\it Gaia} Data Processing and Analysis Consortium (DPAC,\url{https://www.cosmos.esa.int/web/gaia/dpac/consortium}). Funding for the DPAC has been provided by national institutions, in particular the institutions participating in the {\it Gaia} Multilateral Agreement.
% \end{acknowledgments}

\bibliography{paper}{}
\bibliographystyle{aasjournal}

\end{document}